\documentclass[twocolumn,aps,prl]{revtex4}

\usepackage{graphicx}
\begin{document}

\title{Light diffusion and localization in 3D nonlinear disordered media}

\author{C. Conti$^{1,2}$, L. Angelani$^3$,  G. Ruocco$^{2,4}$}

\affiliation{
$^1$Centro studi e ricerche ``Enrico Fermi,'' Via Panisperna 89/A,
I-00184, Roma, Italy   \\
$^2$Research center Soft INFM-CNR, c/o Universit\`a di
Roma ``La Sapienza,'' I-00185, Roma, Italy \\
$^3$Research center SMC INFM-CNR, c/o Universit\`a
di Roma ``La Sapienza,'' I-00185, Roma, Italy \\
$^4$Dipartimento di Fisica, Universit\`a di Roma ``La Sapienza,''
I-00185, Roma, Italy }

\date{\today}

\begin{abstract}
Using a 3D Finite-Difference Time-Domain parallel code, we report on the linear
and nonlinear propagation of light pulses in a disordered assembly of scatterers, whose
spatial distribution is generated by a Molecular Dynamics code;
refractive index dispersion is also taken into account.
We calculate the static and dynamical diffusion constant of light, while considering a 
pulsed excitation. Our results are in quantitative agreement with reported
experiments, also furnishing evidence of a non-exponential decay of the transmitted 
pulse in the linear regime and in the presence of localized modes.
By using an high power excitation, we numerically demonstrate the ``modulational 
instability random laser'':
at high peak input powers energy is transferred to localized states from the input pulse,
via third-order nonlinearity and optical parametric amplification, and this process 
is signed by a power-dependent  non-exponential time-decay of the transmitted pulse.
\end{abstract}

\maketitle
Electromagnetic wave localization and diffusion in random media have been recently considered by several authors.
Experimental studies reported on various evidences of light localization, 
including the existence of a non-exponential long-time tail 
of the transmitted intensity \cite{Storzer06, Chabanov04, Chabanov01,Chabanov03, Vellekoop05, Johnson03, Wiersma97, Reil05}, 
and these processes were considered in numerous theoretical and numerical papers 
(as e.g. in \cite{RevModPhys.71.313, Skipetrov06,Vanneste05, Cheung05, Markos05,Tseng04,Tseng05,Tseng05a}). 
In a purely diffusive regime, pulsed light impinging on a disordered sample is rapidly dispersed while
propagating; this implies that nonlinear effects are typically negligible.
This holds true as far as localization processes do not get involved.
For example, in random lasers (for a recent review see \cite{Cao05}), localized or extended modes \cite{Mujumdar04}  
are excited in a disordered medium by means of an optically active material;
in this case, the large resonant nonlinear susceptibility and the fact that long living
(high Q) modes are excited foster a variety of nonlinear phenomena (as e.g. in \cite{Liu03,Deych05, Angelani06}).
For non-resonant ultra-fast nonlinearities (as $\chi^{(3)}$ processes of electronic origin \cite{BoydBook}),
it is in principle possible to use high-peak intensity femtosecond laser pulses to induce nonlinear effects,
even in the presence of strong light diffusion.
However, the (numerical and theoretical) analysis is enormously complicated 
by the need to include various effects as disorder, nonlinearity, 
material dispersion and to take into account a 3D environment.
Even in the absence of nonlinearity, no ``ab initio'' numerical investigation 
of 3D Maxwell equations for femtosecond pulses in disordered media has been reported, to the best of our knowledge. 
Some theoretical investigations considered 
the role of the Kerr effect (i.e. an intensity dependent refractive index) for a monochromatic excitation \cite{Spivak00,Skipetrov00, Skipetrov01,Skipetrov03};
however ultra-fast $\chi^{(3)}$ optical nonlinearity may also provide other mechanisms. 
We specifically consider the gain
of non-resonant origin, i.e. the optical parametric oscillation (OPO).
OPO, in isotropic media in the absence of disorder, 
has been recently considered with reference to various devices \cite{Conti04, Foster06,Kippenberg04},  
while in the fiber-optics community it is known as the ``modulational instability laser'' \cite{Nakazawa89}. 
Here we report on the  extensive 3D+1 numerical analysis of linear diffusion and nonlinear amplification processes
in a random dispersive nonlinear medium. We use a Molecular Dynamics (MD) code for generating a three dimensional distribution 
of spherical scatterers and a Finite-Difference Time-Domain (FDTD) parallel algorithm for simulating light propagation. 
The MD-FDTD approach not only reproduces the diffusive regime in agreement with reported experimental results in the linear regime,
but also provides clear indications of the random OPO process, with analogies to random lasers.

An ``ab initio'' investigation of the nonlinear light propagation in 3D disordered media, 
requires a realistic distribution of a colloidal medium and of its
nonlinear dielectric response.
We model the disordered system as an assembly of poly-dispersed particles in air. 
The MD code generates the particle configuration;
specifically, we considered a 50:50 mixture of spheres with diameter ratio $1.2$, 
interacting by a $100-200$ Lennard Jones potential \cite{Angelani04}.
The considered sample is composed by $1000$ scatterers; 
once determined the particle positions by the MD code, each of them is taken as filled by a dispersive 
optically nonlinear medium, which is modeled as described below. The particle dimensions are 
chosen in order to simulate typical samples used in light localization and diffusion experiments,
as e.g. in \cite{Vellekoop05,Storzer06}.
We considered diameters for the biggest particles in the mixture ranging 
from $200$ nm to $400$ nm; correspondingly the volume filling fraction $\phi$ for
our samples, with volume $L^3$ ($L=4$ $\mu$m, of the order of the values considered in \cite{Johnson03}), varies in the interval $\phi\in[0.1 , 0.6]$.
The optical linear response of the particle medium is chosen as the best approximation of that of 
$TiO_2$, used in most of the reported experiments, including material dispersion.
The latter is modeled by a single-pole Lorentz medium, reproducing the
 data for $TiO_2$ reported in \cite{DeVore51}. 
Material absorption is also included providing an absorption coefficient
$\kappa_i=10$ mm$^{-1}$ at $\lambda=514$ nm.
The 3D Maxwell equations, solved by the parallel FDTD \cite{TafloveBook} algorithm, are written as 
\begin{equation}
\begin{array}{l}
\label{maxwell}
\nabla\times {\bf E}=-\mu_0 \partial_t {\bf H} \hspace{1cm} \nabla\times {\bf H}=\epsilon_0 \epsilon_{\infty}\partial_t {\bf E}+\partial_t {\bf P}\\
\partial_t^2 {\bf P} +2\gamma_0 \partial_t {\bf P} +\omega_0^2 f(P) {\bf P} =\epsilon_0 (\epsilon_s-\epsilon_{\infty}) \omega_0^2 {\bf E}
\end{array}
\end{equation}
The non-resonant nonlinear response is determined by 
the nonlinear Lorentz oscillator (see e.g. \cite{BoydBook}).
As $f(P)=1$,  eqs.(\ref{maxwell}) model the linear dispersive medium $Ti O_2$;
by appropriately choosing $f(P)$, the system displays a isotropic ultra-fast nonlinearity, as previously described in \cite{Conti04},
including the non-resonant $\chi^{(3)}$ susceptibility.
In the present case, the associated Kerr effect has a nonlinear refractive index $\Delta n= n_2 I$, with $I$ the
optical intensity and $n_2 \cong 10^{-18}$m$^2$W$^{-1}$.
The parameters in $TiO_2$ particles are given by (MKS units)  : 
$\epsilon_s=5.9130$, $\epsilon_\infty=2.8731$ , $\omega_0=6.6473\times 10^{15}$, $\gamma_0=8.9168\times 10^{11}$ and 
$f(P)=[1+(P/P_0)^2]^{3/2}$ with $P_0^{-2}=3.0$;
out of the particles, $P=0$ and $\epsilon_s=\epsilon_\infty=1$ (vacuum).

The code is parallelized using the ``MPI'' protocol,
and ``UPML'' absorbing boundary condition are used in the three spatial directions \cite{TafloveBook}.
The used discretization is $\Delta x=\Delta y=\Delta z\cong 30$ nm and $\Delta t\cong0.02$ fs.
The $z-$propagating input beam is taken with a Gaussian TEM$_{00}$  
linearly $y-$polarized profile, at the input face of the disordered medium its waist is 
$w_0=1\ \mu$m. The input pulse temporal profile is Gaussian, with duration $t_0=100$ fs and carrier wavelength $\lambda=520$ nm.
The numerical results have been verified a number of times by halving the spatio-temporal discretization. 
The validity of our approach has been confirmed in a variety of studies, as e.g. in \cite{Conti04,Conti04b}.

The first numerical simulations were aimed to investigate light diffusion 
when the nonlinear effects can be taken as negligible (low power),
for a fixed sample spatial extension. We considered increasing values for the diameter 
of the spheres in the mixture; in the following $\sigma$ is the corresponding value of the
biggest particles in the considered 50:50 mixture (for the others, the diameter is $\sigma/1.2$).
This corresponds to increase the average density of the medium, as well as the volume packing fraction. 
In order to unveil the onset of diffusion, we monitored the time-dependent output intensity,
which is expected to develop an exponential tail in the diffusive regime.
This is obtained in two ways: i) we considered 
the electric field $E_y$ at one point in the middle of the output face of the sample
(see figure 1),
and we calculated the ``local intensity'' by squaring the field and filtering with a low 
pass filter to remove the optical carrier (thus mimicking a photodiode); ii) we considered the total transmission
from the output face of the sample as the calculated $z-$component of the Poynting vector, integrated with respect to
the transverse $x,y$ coordinates.

In figure 1 we show the results for the local intensity, while increasing the volume filling fraction $\phi$.
In order to compare the trailing edge of the transmitted pulses,
for each signal we rescaled the peak value to the unity and shifted the temporal axis to make all the
peaks for any $\sigma$ coincident.
At high packing fraction the onset of an exponential trend is clearly visible.
Note the residual oscillations, that are due to the fact that light
is collected in a specific point at the output face. 
A typical far-field speckle pattern calculated from our simulation is shown in the inset of the bottom panel, and 
obtained by storing the $E_y$ component of the electric field in the output section,
Fourier transforming with respect to the transverse coordinates, and averaging over time
(the pattern is calculated for a continuous wave excitation at $\lambda=520$ nm).

The intensity modulation in figure 1 disappears when considering the overall transmission 
from the output facet, which provides an average information over all the speckles
and is shown in figure \ref{figpoynting}.
The instantaneous flux for the z-component of the Poynting vector is calculated and 
then filtered to remove the reactive component.
The appearance of an exponential tail in the transmission denotes the transition to the diffusive regime.
The transmitted signal is $I(t)\propto \exp[-\pi^2 D(t) t/L^2]$, with an instantaneous diffusion coefficient 
$D(t)\rightarrow D$ as $t\rightarrow\infty$
and a sample length $L$.
When the packing fraction increases, the tail gets longer due to the reduced diffusivity $D$.
The latter quantity is shown in Fig. \ref{figpoynting}(b) vs $\phi$. The results are in agreement
with experimentally determined values, as reported in literature (e.g. in \cite{ Johnson03, Vellekoop05,Storzer06}). 
Note that, as expected, $D$ decreases as the volume packing fraction increases,
and at high $\phi$ the diffusivity tends to an asymptotic value (for non overlapping particles).
The inset in Fig. \ref{figpoynting}(b) shows the calculated instantaneous diffusivity for two values of 
$\sigma$, which reproduces the reported trends, including some temporal oscillations \cite{Chabanov03, Storzer06}. 

Next, we consider the diffusive regime for a fixed $\sigma=400$ nm ($\phi\cong0.6$), while increasing the input peak power $P_{in}$.
In figure \ref{fignl}(a), we show the local intensity for two input peak powers; the onset of a non-exponential tail is 
evident, as also shown in the inset reporting the intensity auto-correlation function.
In the panel (b) we show the transmitted intensity (integrated Poynting vector) for various input powers.
The inset shows the instantaneous diffusivity that drops at high powers and long times, when a threshold
value for the input peak power is reached.
These results foster the interpretation of the data in terms of a light-induced localization transition,
strongly resembling that controlled e.g.
by the packing fraction and experimentally observed (as e.g. in \cite{Storzer06}).
However, in the present case, the transition is controlled
by the input power, and hence it has a nonlinear origin.
It will be shown below that it can be explained as the excitation of
a ``modulational instability random laser'' (or random OPO).

Indeed, to unveil the underlying  mechanism, we resorted to the spectral analysis of the
electric field in various points inside the random medium;
figure \ref{figspectrum} summarizes the typical result.
In panel (a) we show the calculated frequency spectra of the electric field $E_y$ in the
random structure, for different input peak powers.
The data are displayed in logarithmic scale and vertically shifted by an arbitrary amount for the
sake of clearness.
At low power the spectrum of the input pulse is reproduced, with the addition of small ripples
(due to the vertical scale in figure  \ref{figspectrum}a)  
in correspondence of the existing high-Q modes. 
As the power increases, the 
bandwidth gets larger and regions with depression and enhancement of the spectrum appear.
At sufficiently high power, a wide band excitation is found.
This behavior can be explained by recalling that in $\chi^{(3)}$ media energy can be transferred 
from various harmonics to others through parametric amplification.
We stress that the nonlinear effects are weakly pertubative, indeed we consider power level just above the
OPO threshold and the nonlinear index perturbation, due to the Kerr effect, is of the order of $0.001$.
Thus the localization process in figure \ref{fignl} is explained by  assuming that
energy is nonlinearly transferred to localized, long-living states. 

In panel (b) we show the time evolution of the spectrum (i.e. the spectrogram calculated by
Fourier transforming windowed parts of the temporal signal) at $P_{in}=3$ kW. 
In the initial interval $t<1$ ps, energy is transferred
to a wide band around the input pulse spectrum. However, at large times only the long-lifetime (high Q-factor) modes 
survive and the spectrum is characterized by several peaks in the region
around $\lambda\cong300$ nm ($\nu\cong 0.9\times 10^{15}$ Hz).
In order to sustain the fact that localized modes put into oscillation, we show in the inset
of figure 4(c) a snapshot of the electric field $E_y$ in the sample, taken at $P=2.5$ kW
and $t\cong4$ ps which clearly show the occurrence of localized excitations.

\begin{figure}
\includegraphics[width=8.3cm]{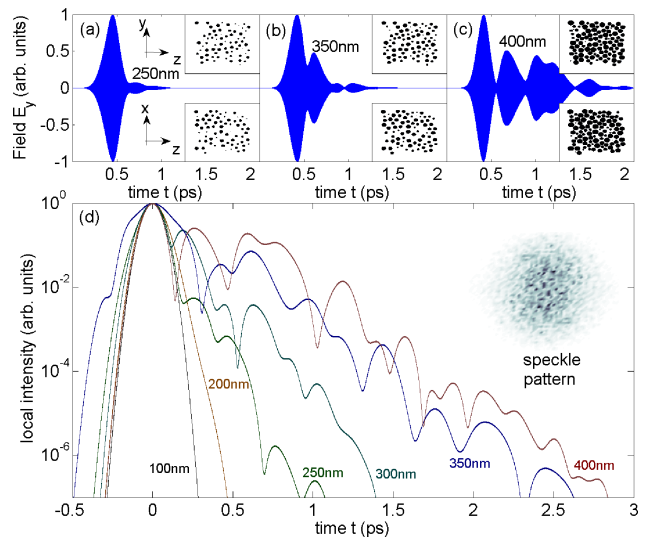}
\caption{
(Color online) Top panels (a-c): electric field in the middle of the output face of the sample at low 
input peak power ($P_{in}=1$ nW), for three different particle diameters ($\sigma=250,350,400$ nm).
The insets show the particle distribution in
the yz (top) and xz (bottom) middle section of the sample.
Bottom panel (d): local intensity in logarithmic scale. The inset shows 
the calculated speckle pattern.
\label{figlinear}}
\end{figure}
\begin{figure}
\includegraphics[width=8.3cm]{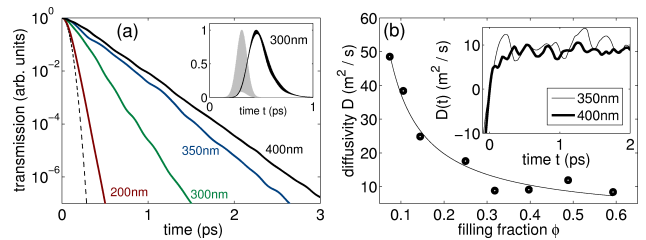}
\caption{
(Color online) (a) Tail of the transmitted pulse (the flux of the z-component of the Poynting vector at the output face) for various particle diameters. The inset shows the transmitted pulse in linear scale, the gray region is the input pulse.
(b) Calculated diffusion constant for various filling fractions. The continuous line
is a best-fit by an inverse power law.
The inset shows the dynamical diffusivity for two values of the particle diameters.
\label{figpoynting}}
\end{figure}

In addition, we went back to the linear regime to analyze the localized
modes are supported in our virtual samples. 
First, we repeated the pulse transmission experiment at low power $P=1$ nW and central wavelength
$\lambda=300$ nm ($\sigma=400$ nm) and Figure 4(c) clearly shows an non-exponential tail
at $t\cong 2.5$ ps, which is not present for $\lambda=521$ nm (figure 2a) also for $t>3$ ps (not shown).
Then, we recall that light localization
is typically expected in proximity of depressions of the density of states (DOS).
For the case of weakly disordered periodical structures, these are obtained inside the photonic band gaps \cite{John87},
while, for random dispersions of particles, the DOS (which
is numerically calculated by shining the sample by a ultra-wide band single-cycle pulse and spectrally analyzing 
the transmitted pulse) displays some depressions (pseudo-gaps) in proximity of the Mie resonances (see e.g. \cite{Vanneste05, Chabanov01}).
In our case, the distribution of states displays various depressions with the most pronounced around
$0.9 \times 10^{15}$ Hz, as shown in Fig. \ref{figspectrum}(d).
The fact that localized modes do exist in proximity of the pseudo-gaps, is also confirmed by the
plot of the calculated Q-factor vs frequency (bold line in Fig. \ref{figspectrum}(d)), realized by using an harmonic inversion 
library \cite{Mandelshtam01}.
Which of the localized modes are effectively excited 
will depend not only on the Q-factors, but also on the spatial overlaps with the pump modes and with
the spatial distribution of the nonlinear susceptibility (the $g$ coefficient discussed below). 
The localized modes are a reminiscence of the Mie resonances of the isolated spheres, which are strongly affected
by the presence of adjacent resonators in a disordered configuration, in analogy to what happen for periodical
mono-dispersed photonic crystal, investigated e.g. in \cite{Vandenbem05}.

\begin{figure}
\includegraphics[width=8.3cm]{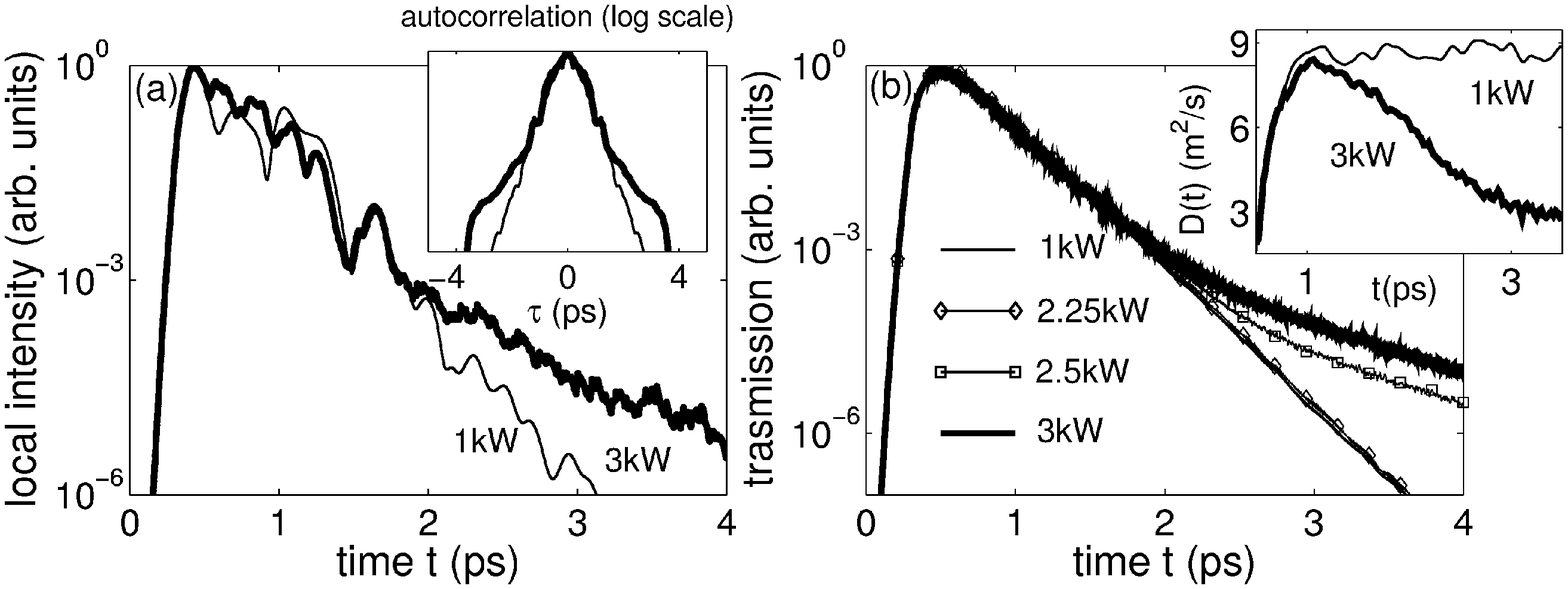}
\caption{ 
(a) Time dynamics of the local intensity for two different input peak powers; the
inset shows the corresponding auto-correlation functions versus the delay $\tau$;
(b) transmitted signal for various input peak powers; the inset shows the calculated 
instantaneous diffusivity.
\label{fignl}}
\end{figure}
\begin{figure}
\includegraphics[width=8.3cm]{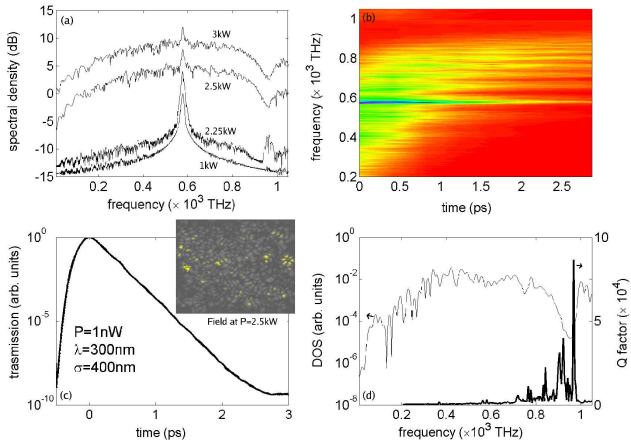}
\caption{(Color online) 
(a) Spectral emission for increasing peak power;
(b) spectrogram displaying the time evolution of the spectra for $P=3$ kW;
(c) transmission of a low power pulse at $\lambda=300$ nm and $\sigma=400$ nm displaying 
a not-exponential tails; the inset shows the field distribution for $\sigma=400$ nm 
at high power ($P=2.5$ kW) at $t=4$ ps in the xz middle section of the sample;
(d) density of states and Q-factor distribution for $\sigma=400$ nm.
\label{figspectrum}}
\end{figure}

Before concluding we show that the basic features of the reported results can 
be reproduced by a simple theoretical model, based on the coupled mode theory in the
time domain (CMT) \cite{HausBook}. Indeed the small-signal parametric gain for the four-waves interaction
$\omega_0+\omega_0=(\omega_0+\Omega)+(\omega_0-\Omega)$, can be determined by introducing the 
complex amplitudes for the pump mode $a_0(t)=a(\omega_0,t)$ and for the generated modes 
$a_\pm(t)=a(\omega_0\pm \Omega,t)$. The complex amplitudes are normalized in a way such that $|a(\omega,t)|^2$ is
the energy stored in the mode at angular frequency $\omega$, and we denote with $\tau(\omega)$ the lifetime,
 and $Q(\omega)=\omega \tau(\omega)/2$ the corresponding Q-factor. 
In the un-depleted pump approximation the relevant CMT equations are written
in compact and obvious notation as
\begin{equation}
\displaystyle\frac{d a_\pm}{dt}=-\frac{1}{ \tau_\pm} a_\pm+ i g a_0^2 a_\mp^*\text{.}
\label{CMT}
\end{equation}
The coefficient $g$ is the relevant 3D spatial overlap between the mode electric field distributions and the nonlinear third-order susceptibility. 
Looking for solutions in the form 
$a_\pm=A_\pm \exp(\alpha t)$, the real-valued gain $\alpha$ is given by
\begin{equation}
\alpha=-\left(\frac{1}{2\tau_+}+\frac{1}{2\tau_-}\right)+\left[ \left( 
\frac{1}{2\tau_+}-\frac{1}{2\tau_-} \right)^2 +|g a_0^2|^2 \right]^{1/2} \text{.}
\end{equation}
It is readily seen 
that a condition for having amplification ($\alpha>0$) is given by 
$|g a_0^2|^2> 1/(\tau_+ \tau_-)= \omega_+ \omega_+ /4 Q_+ Q_-$.
Assuming that for all the involved modes $\omega\cong \omega_0$ and the Q-factor is almost the same,
with the exception of those localized that display higher values, it is concluded that the localized modes 
(those with not negligible overlap $g$ with the pump modes)
are put into oscillations at lower pump energies (see line at $2.25$ kW in figure \ref{figspectrum}(a)). 
Additionally the asymmetry, at low power, in the spectrum (see \ref{figspectrum}a) can be explained by observing from Eq. (\ref{CMT}) that $|A_+/A_-|>1$
if $Q_+>Q_-$. At higher powers most of the modes are put into oscillation and the spectrum mimics the DOS.

In conclusion, the considered disordered, nonlinear and dispersive samples support
localized modes in proximity of Mie resonances. These can be excited by high-intensity
pulses through optical parametric amplification.
The overall process can be interpreted as the excitation of a ``modulational instability random laser''.
The main difference with standard random lasers is
 given by the fact that an ultra-fast optical process provides the required gain
and does not need doping of the sample by an active media.
These results can be generalized to other nonlinear amplification processes, 
like the Raman effect, and can find application in all-optical storage of light,
ultra-fast laser-tissue interaction, soft-matter and granular systems.
We thank S. Trillo and D. Wiersma for fruitful discussions.
We acknowledge support from the INFM-CINECA initiative for parallel computing.


\begin{thebibliography}{38}
\expandafter\ifx\csname natexlab\endcsname\relax\def\natexlab#1{#1}\fi
\expandafter\ifx\csname bibnamefont\endcsname\relax
  \def\bibnamefont#1{#1}\fi
\expandafter\ifx\csname bibfnamefont\endcsname\relax
  \def\bibfnamefont#1{#1}\fi
\expandafter\ifx\csname citenamefont\endcsname\relax
  \def\citenamefont#1{#1}\fi
\expandafter\ifx\csname url\endcsname\relax
  \def\url#1{\texttt{#1}}\fi
\expandafter\ifx\csname urlprefix\endcsname\relax\def\urlprefix{URL }\fi
\providecommand{\bibinfo}[2]{#2}
\providecommand{\eprint}[2][]{\url{#2}}

\bibitem[{\citenamefont{Storzer et~al.}(2006)\citenamefont{Storzer, Gross,
  Aegerter, and Maret}}]{Storzer06}
\bibinfo{author}{\bibfnamefont{M.}~\bibnamefont{Storzer}},
  \bibinfo{author}{\bibfnamefont{P.}~\bibnamefont{Gross}},
  \bibinfo{author}{\bibfnamefont{C.~M.} \bibnamefont{Aegerter}},
  \bibnamefont{and} \bibinfo{author}{\bibfnamefont{G.}~\bibnamefont{Maret}},
  \bibinfo{journal}{\prl} \textbf{\bibinfo{volume}{96}},
  \bibinfo{pages}{063904} (\bibinfo{year}{2006}).

\bibitem[{\citenamefont{Chabanov et~al.}(2004)\citenamefont{Chabanov, Hu, and
  Genack}}]{Chabanov04}
\bibinfo{author}{\bibfnamefont{A.~A.} \bibnamefont{Chabanov}},
  \bibinfo{author}{\bibfnamefont{B.}~\bibnamefont{Hu}}, \bibnamefont{and}
  \bibinfo{author}{\bibfnamefont{A.~Z.} \bibnamefont{Genack}},
  \bibinfo{journal}{\prl} \textbf{\bibinfo{volume}{93}},
  \bibinfo{pages}{123901} (\bibinfo{year}{2004}).

\bibitem[{\citenamefont{Chabanov and Genack}(2001)}]{Chabanov01}
\bibinfo{author}{\bibfnamefont{A.~A.} \bibnamefont{Chabanov}} \bibnamefont{and}
  \bibinfo{author}{\bibfnamefont{A.~Z.} \bibnamefont{Genack}},
  \bibinfo{journal}{\prl} \textbf{\bibinfo{volume}{87}},
  \bibinfo{pages}{153901} (\bibinfo{year}{2001}).

\bibitem[{\citenamefont{Chabanov et~al.}(2003)\citenamefont{Chabanov, Zhang,
  and Genack}}]{Chabanov03}
\bibinfo{author}{\bibfnamefont{A.~A.} \bibnamefont{Chabanov}},
  \bibinfo{author}{\bibfnamefont{Z.~Q.} \bibnamefont{Zhang}}, \bibnamefont{and}
  \bibinfo{author}{\bibfnamefont{A.~Z.} \bibnamefont{Genack}},
  \bibinfo{journal}{\prl} \textbf{\bibinfo{volume}{90}},
  \bibinfo{pages}{203903} (\bibinfo{year}{2003}).

\bibitem[{\citenamefont{Vellekoop et~al.}(2005)\citenamefont{Vellekoop, Lodahl,
  and Lagendijk}}]{Vellekoop05}
\bibinfo{author}{\bibfnamefont{I.~M.} \bibnamefont{Vellekoop}},
  \bibinfo{author}{\bibfnamefont{P.}~\bibnamefont{Lodahl}}, \bibnamefont{and}
  \bibinfo{author}{\bibfnamefont{A.}~\bibnamefont{Lagendijk}},
  \bibinfo{journal}{\pre} \textbf{\bibinfo{volume}{71}},
  \bibinfo{pages}{056604} (\bibinfo{year}{2005}).

\bibitem[{\citenamefont{Johnson et~al.}(2003)\citenamefont{Johnson, Imhof,
  Bret, Rivas, and Lagendijk}}]{Johnson03}
\bibinfo{author}{\bibfnamefont{P.~M.} \bibnamefont{Johnson}},
  \bibinfo{author}{\bibfnamefont{A.}~\bibnamefont{Imhof}},
  \bibinfo{author}{\bibfnamefont{B.~P.~J.} \bibnamefont{Bret}},
  \bibinfo{author}{\bibfnamefont{J.~G.} \bibnamefont{Rivas}}, \bibnamefont{and}
  \bibinfo{author}{\bibfnamefont{A.}~\bibnamefont{Lagendijk}},
  \bibinfo{journal}{\pre} \textbf{\bibinfo{volume}{68}},
  \bibinfo{pages}{016604} (\bibinfo{year}{2003}).

\bibitem[{\citenamefont{Wiersma et~al.}(1997)\citenamefont{Wiersma, Bartolini,
  Lagendijk, and Righini}}]{Wiersma97}
\bibinfo{author}{\bibfnamefont{D.~S.} \bibnamefont{Wiersma}},
  \bibinfo{author}{\bibfnamefont{P.}~\bibnamefont{Bartolini}},
  \bibinfo{author}{\bibfnamefont{A.}~\bibnamefont{Lagendijk}},
  \bibnamefont{and} \bibinfo{author}{\bibfnamefont{R.}~\bibnamefont{Righini}},
  \bibinfo{journal}{Nature} \textbf{\bibinfo{volume}{390}},
  \bibinfo{pages}{671} (\bibinfo{year}{1997}).

\bibitem[{\citenamefont{Reil and Thomas}(2005)}]{Reil05}
\bibinfo{author}{\bibfnamefont{F.}~\bibnamefont{Reil}} \bibnamefont{and}
  \bibinfo{author}{\bibfnamefont{J.~E.} \bibnamefont{Thomas}},
  \bibinfo{journal}{\prl} \textbf{\bibinfo{volume}{95}},
  \bibinfo{pages}{143903} (\bibinfo{year}{2005}).

\bibitem[{\citenamefont{van Rossum and
  Nieuwenhuizen}(1999)}]{RevModPhys.71.313}
\bibinfo{author}{\bibfnamefont{M.~C.~W.} \bibnamefont{van Rossum}}
  \bibnamefont{and} \bibinfo{author}{\bibfnamefont{T.~M.}
  \bibnamefont{Nieuwenhuizen}}, \bibinfo{journal}{Rev. Mod. Phys.}
  \textbf{\bibinfo{volume}{71}}, \bibinfo{pages}{313} (\bibinfo{year}{1999}).

\bibitem[{\citenamefont{Skipetrov and van Tiggelen}(2006)}]{Skipetrov06}
\bibinfo{author}{\bibfnamefont{S.~E.} \bibnamefont{Skipetrov}}
  \bibnamefont{and} \bibinfo{author}{\bibfnamefont{B.~A.} \bibnamefont{van
  Tiggelen}}, \bibinfo{journal}{\prl} \textbf{\bibinfo{volume}{96}},
  \bibinfo{pages}{043902} (\bibinfo{year}{2006}).

\bibitem[{\citenamefont{Vanneste and Sebbah}(2005)}]{Vanneste05}
\bibinfo{author}{\bibfnamefont{C.}~\bibnamefont{Vanneste}} \bibnamefont{and}
  \bibinfo{author}{\bibfnamefont{P.}~\bibnamefont{Sebbah}},
  \bibinfo{journal}{\pre} \textbf{\bibinfo{volume}{71}},
  \bibinfo{pages}{026612} (\bibinfo{year}{2005}).

\bibitem[{\citenamefont{Cheung and Zhang}(2005)}]{Cheung05}
\bibinfo{author}{\bibfnamefont{S.~K.} \bibnamefont{Cheung}} \bibnamefont{and}
  \bibinfo{author}{\bibfnamefont{Z.~Q.} \bibnamefont{Zhang}},
  \bibinfo{journal}{\prb} \textbf{\bibinfo{volume}{72}},
  \bibinfo{pages}{235102} (\bibinfo{year}{2005}).

\bibitem[{\citenamefont{Markos and Soukoulis}(2005)}]{Markos05}
\bibinfo{author}{\bibfnamefont{P.}~\bibnamefont{Markos}} \bibnamefont{and}
  \bibinfo{author}{\bibfnamefont{C.~M.} \bibnamefont{Soukoulis}},
  \bibinfo{journal}{\prb} \textbf{\bibinfo{volume}{71}},
  \bibinfo{pages}{054201} (\bibinfo{year}{2005}).

\bibitem[{\citenamefont{Tseng et~al.}(2004)\citenamefont{Tseng, Greene,
  Taflove, Maitland, Backman, V., and Walsh~Jr.}}]{Tseng04}
\bibinfo{author}{\bibfnamefont{S.~H.} \bibnamefont{Tseng}},
  \bibinfo{author}{\bibfnamefont{J.~H.} \bibnamefont{Greene}},
  \bibinfo{author}{\bibfnamefont{A.}~\bibnamefont{Taflove}},
  \bibinfo{author}{\bibfnamefont{D.}~\bibnamefont{Maitland}},
  \bibinfo{author}{\bibfnamefont{V.}~\bibnamefont{Backman}},
  \bibinfo{author}{\bibnamefont{V.}}, \bibnamefont{and}
  \bibinfo{author}{\bibfnamefont{J.~T.} \bibnamefont{Walsh~Jr.}},
  \bibinfo{journal}{\ol} \textbf{\bibinfo{volume}{29}}, \bibinfo{pages}{1393}
  (\bibinfo{year}{2004}).

\bibitem[{\citenamefont{Tseng et~al.}(2005{\natexlab{a}})\citenamefont{Tseng,
  Kim, Taflove, Maitland, Backman, and Walsh~Jr.}}]{Tseng05}
\bibinfo{author}{\bibfnamefont{S.~H.} \bibnamefont{Tseng}},
  \bibinfo{author}{\bibfnamefont{Y.~L.} \bibnamefont{Kim}},
  \bibinfo{author}{\bibfnamefont{A.}~\bibnamefont{Taflove}},
  \bibinfo{author}{\bibfnamefont{D.}~\bibnamefont{Maitland}},
  \bibinfo{author}{\bibfnamefont{V.}~\bibnamefont{Backman}}, \bibnamefont{and}
  \bibinfo{author}{\bibfnamefont{J.~T.} \bibnamefont{Walsh~Jr.}},
  \bibinfo{journal}{Opt. Express} \textbf{\bibinfo{volume}{13}},
  \bibinfo{pages}{3666} (\bibinfo{year}{2005}{\natexlab{a}}).

\bibitem[{\citenamefont{Tseng et~al.}(2005{\natexlab{b}})\citenamefont{Tseng,
  Taflove, Maitland, Backman, and Walsh~Jr.}}]{Tseng05a}
\bibinfo{author}{\bibfnamefont{S.~H.} \bibnamefont{Tseng}},
  \bibinfo{author}{\bibfnamefont{A.}~\bibnamefont{Taflove}},
  \bibinfo{author}{\bibfnamefont{D.}~\bibnamefont{Maitland}},
  \bibinfo{author}{\bibfnamefont{V.}~\bibnamefont{Backman}}, \bibnamefont{and}
  \bibinfo{author}{\bibfnamefont{J.~T.} \bibnamefont{Walsh~Jr.}},
  \bibinfo{journal}{Opt. Express} \textbf{\bibinfo{volume}{13}},
  \bibinfo{pages}{6127} (\bibinfo{year}{2005}{\natexlab{b}}).

\bibitem[{\citenamefont{Cao}(2005)}]{Cao05}
\bibinfo{author}{\bibfnamefont{H.}~\bibnamefont{Cao}}, \bibinfo{journal}{J.
  Phys. A. : Math. Gen.} \textbf{\bibinfo{volume}{38}}, \bibinfo{pages}{10497}
  (\bibinfo{year}{2005}).

\bibitem[{\citenamefont{Mujumdar et~al.}(2004)\citenamefont{Mujumdar, Ricci,
  Torre, and Wiersma}}]{Mujumdar04}
\bibinfo{author}{\bibfnamefont{S.}~\bibnamefont{Mujumdar}},
  \bibinfo{author}{\bibfnamefont{M.}~\bibnamefont{Ricci}},
  \bibinfo{author}{\bibfnamefont{R.}~\bibnamefont{Torre}}, \bibnamefont{and}
  \bibinfo{author}{\bibfnamefont{D.~S.} \bibnamefont{Wiersma}},
  \bibinfo{journal}{\prl} \textbf{\bibinfo{volume}{93}},
  \bibinfo{pages}{053903} (\bibinfo{year}{2004}).

\bibitem[{\citenamefont{Liu et~al.}(2003)\citenamefont{Liu, Yamilov, Ling, Xu,
  and Cao}}]{Liu03}
\bibinfo{author}{\bibfnamefont{B.}~\bibnamefont{Liu}},
  \bibinfo{author}{\bibfnamefont{A.}~\bibnamefont{Yamilov}},
  \bibinfo{author}{\bibfnamefont{Y.}~\bibnamefont{Ling}},
  \bibinfo{author}{\bibfnamefont{J.~Y.} \bibnamefont{Xu}}, \bibnamefont{and}
  \bibinfo{author}{\bibfnamefont{H.}~\bibnamefont{Cao}},
  \bibinfo{journal}{\prl} \textbf{\bibinfo{volume}{91}},
  \bibinfo{pages}{063903} (\bibinfo{year}{2003}).

\bibitem[{\citenamefont{Deych}(2005)}]{Deych05}
\bibinfo{author}{\bibfnamefont{L.~I.} \bibnamefont{Deych}},
  \bibinfo{journal}{\prl} \textbf{\bibinfo{volume}{95}},
  \bibinfo{pages}{043902} (\bibinfo{year}{2005}).

\bibitem[{\citenamefont{Angelani et~al.}(2006)\citenamefont{Angelani, Conti,
  Ruocco, and Zamponi}}]{Angelani06}
\bibinfo{author}{\bibfnamefont{L.}~\bibnamefont{Angelani}},
  \bibinfo{author}{\bibfnamefont{C.}~\bibnamefont{Conti}},
  \bibinfo{author}{\bibfnamefont{G.}~\bibnamefont{Ruocco}}, \bibnamefont{and}
  \bibinfo{author}{\bibfnamefont{F.}~\bibnamefont{Zamponi}},
  \bibinfo{journal}{\prl} \textbf{\bibinfo{volume}{96}},
  \bibinfo{pages}{065702} (\bibinfo{year}{2006}).

\bibitem[{\citenamefont{Boyd}(2002)}]{BoydBook}
\bibinfo{author}{\bibfnamefont{R.~W.} \bibnamefont{Boyd}},
  \emph{\bibinfo{title}{Nonlinear Optics}} (\bibinfo{publisher}{Academic
  Press}, \bibinfo{address}{New York}, \bibinfo{year}{2002}),
  \bibinfo{edition}{2nd} ed.

\bibitem[{\citenamefont{Spivak and Zyuzin}(2000)}]{Spivak00}
\bibinfo{author}{\bibfnamefont{B.}~\bibnamefont{Spivak}} \bibnamefont{and}
  \bibinfo{author}{\bibfnamefont{A.}~\bibnamefont{Zyuzin}},
  \bibinfo{journal}{\prl} \textbf{\bibinfo{volume}{84}}, \bibinfo{pages}{1970}
  (\bibinfo{year}{2000}).

\bibitem[{\citenamefont{Skipetrov and Maynard}(2000)}]{Skipetrov00}
\bibinfo{author}{\bibfnamefont{S.~E.} \bibnamefont{Skipetrov}}
  \bibnamefont{and} \bibinfo{author}{\bibfnamefont{R.}~\bibnamefont{Maynard}},
  \bibinfo{journal}{\prl} \textbf{\bibinfo{volume}{85}}, \bibinfo{pages}{736}
  (\bibinfo{year}{2000}).

\bibitem[{\citenamefont{Skipetrov}(2001)}]{Skipetrov01}
\bibinfo{author}{\bibfnamefont{S.~E.} \bibnamefont{Skipetrov}},
  \bibinfo{journal}{\pre} \textbf{\bibinfo{volume}{63}},
  \bibinfo{pages}{056614} (\bibinfo{year}{2001}).

\bibitem[{\citenamefont{Skipetrov}(2003)}]{Skipetrov03}
\bibinfo{author}{\bibfnamefont{S.~E.} \bibnamefont{Skipetrov}},
  \bibinfo{journal}{\pre} \textbf{\bibinfo{volume}{67}},
  \bibinfo{pages}{016601} (\bibinfo{year}{2003}).

\bibitem[{\citenamefont{Conti et~al.}(2004{\natexlab{a}})\citenamefont{Conti,
  Di~Falco, and Assanto}}]{Conti04}
\bibinfo{author}{\bibfnamefont{C.}~\bibnamefont{Conti}},
  \bibinfo{author}{\bibfnamefont{A.}~\bibnamefont{Di~Falco}}, \bibnamefont{and}
  \bibinfo{author}{\bibfnamefont{G.}~\bibnamefont{Assanto}},
  \bibinfo{journal}{Opt. Express} \textbf{\bibinfo{volume}{12}},
  \bibinfo{pages}{823} (\bibinfo{year}{2004}{\natexlab{a}}),
  \bibinfo{note}{arXiv:physics/0403013}.

\bibitem[{\citenamefont{Foster et~al.}(2006)\citenamefont{Foster, Turner,
  Sharping, Schmidt, Lipson, and Gaeta}}]{Foster06}
\bibinfo{author}{\bibfnamefont{M.~A.} \bibnamefont{Foster}},
  \bibinfo{author}{\bibfnamefont{A.~C.} \bibnamefont{Turner}},
  \bibinfo{author}{\bibfnamefont{J.~E.} \bibnamefont{Sharping}},
  \bibinfo{author}{\bibfnamefont{B.~S.} \bibnamefont{Schmidt}},
  \bibinfo{author}{\bibfnamefont{M.}~\bibnamefont{Lipson}}, \bibnamefont{and}
  \bibinfo{author}{\bibfnamefont{A.~L.} \bibnamefont{Gaeta}},
  \bibinfo{journal}{Nature} \textbf{\bibinfo{volume}{441}},
  \bibinfo{pages}{960} (\bibinfo{year}{2006}).

\bibitem[{\citenamefont{Kippenberg et~al.}(2004)\citenamefont{Kippenberg,
  Spillane, and Vahala}}]{Kippenberg04}
\bibinfo{author}{\bibfnamefont{T.~J.} \bibnamefont{Kippenberg}},
  \bibinfo{author}{\bibfnamefont{S.~M.} \bibnamefont{Spillane}},
  \bibnamefont{and} \bibinfo{author}{\bibfnamefont{K.~J.}
  \bibnamefont{Vahala}}, \bibinfo{journal}{\prl} \textbf{\bibinfo{volume}{93}},
  \bibinfo{pages}{083904} (\bibinfo{year}{2004}).

\bibitem[{\citenamefont{Nakazawa et~al.}(1989)\citenamefont{Nakazawa, Suzuki,
  and Haus}}]{Nakazawa89}
\bibinfo{author}{\bibfnamefont{M.}~\bibnamefont{Nakazawa}},
  \bibinfo{author}{\bibfnamefont{K.}~\bibnamefont{Suzuki}}, \bibnamefont{and}
  \bibinfo{author}{\bibfnamefont{H.~A.} \bibnamefont{Haus}},
  \bibinfo{journal}{{IEEE} J. Quantum Electron.} \textbf{\bibinfo{volume}{25}},
  \bibinfo{pages}{2036} (\bibinfo{year}{1989}).

\bibitem[{\citenamefont{Angelani et~al.}(2005)\citenamefont{Angelani, Foffi,
  Sciortino, and Tartaglia}}]{Angelani04}
\bibinfo{author}{\bibfnamefont{L.}~\bibnamefont{Angelani}},
  \bibinfo{author}{\bibfnamefont{G.}~\bibnamefont{Foffi}},
  \bibinfo{author}{\bibfnamefont{F.}~\bibnamefont{Sciortino}},
  \bibnamefont{and}
  \bibinfo{author}{\bibfnamefont{P.}~\bibnamefont{Tartaglia}},
  \bibinfo{journal}{J. Phys. : Condens. Matter} \textbf{\bibinfo{volume}{17}},
  \bibinfo{pages}{L113} (\bibinfo{year}{2005}), \bibinfo{note}{we used the data
  at temperature T=0.26 in the units of this reference}.

\bibitem[{\citenamefont{DeVore}(1951)}]{DeVore51}
\bibinfo{author}{\bibfnamefont{J.~R.} \bibnamefont{DeVore}},
  \bibinfo{journal}{J. Opt. Soc. Am.} \textbf{\bibinfo{volume}{41}},
  \bibinfo{pages}{416} (\bibinfo{year}{1951}).

\bibitem[{\citenamefont{Taflove and Hagness}(2000)}]{TafloveBook}
\bibinfo{author}{\bibfnamefont{A.}~\bibnamefont{Taflove}} \bibnamefont{and}
  \bibinfo{author}{\bibfnamefont{S.~C.} \bibnamefont{Hagness}},
  \emph{\bibinfo{title}{Computational Electrodynamics: the finite-difference
  time-domain method}} (\bibinfo{publisher}{Artech House},
  \bibinfo{year}{2000}), \bibinfo{edition}{3rd} ed.

\bibitem[{\citenamefont{Conti et~al.}(2004{\natexlab{b}})\citenamefont{Conti,
  Di~Falco, and Assanto}}]{Conti04b}
\bibinfo{author}{\bibfnamefont{C.}~\bibnamefont{Conti}},
  \bibinfo{author}{\bibfnamefont{A.}~\bibnamefont{Di~Falco}}, \bibnamefont{and}
  \bibinfo{author}{\bibfnamefont{G.}~\bibnamefont{Assanto}},
  \bibinfo{journal}{\ol} \textbf{\bibinfo{volume}{29}}, \bibinfo{pages}{2902}
  (\bibinfo{year}{2004}{\natexlab{b}}).

\bibitem[{\citenamefont{John}(1987)}]{John87}
\bibinfo{author}{\bibfnamefont{S.}~\bibnamefont{John}}, \bibinfo{journal}{\prl}
  \textbf{\bibinfo{volume}{58}}, \bibinfo{pages}{2486} (\bibinfo{year}{1987}).

\bibitem[{\citenamefont{Mandelshtam}(2001)}]{Mandelshtam01}
\bibinfo{author}{\bibfnamefont{V.~A.} \bibnamefont{Mandelshtam}},
  \bibinfo{journal}{Progress in NMR Spectroscopy}
  \textbf{\bibinfo{volume}{38}}, \bibinfo{pages}{159} (\bibinfo{year}{2001}),
  \bibinfo{note}{we used the Harminv library by Steven G. Johnson (MIT).}

\bibitem[{\citenamefont{Vandenbem and Vigneron}(2005)}]{Vandenbem05}
\bibinfo{author}{\bibfnamefont{C.}~\bibnamefont{Vandenbem}} \bibnamefont{and}
  \bibinfo{author}{\bibfnamefont{J.~P.} \bibnamefont{Vigneron}},
  \bibinfo{journal}{\josaa} \textbf{\bibinfo{volume}{22}},
  \bibinfo{pages}{1042} (\bibinfo{year}{2005}).

\bibitem[{\citenamefont{Haus}(1984)}]{HausBook}
\bibinfo{author}{\bibfnamefont{H.~A.} \bibnamefont{Haus}},
  \emph{\bibinfo{title}{Waves and Fields in Optoelectronics}}
  (\bibinfo{publisher}{Prentice-Hall}, \bibinfo{address}{Englewood Cliffs, N.
  J.}, \bibinfo{year}{1984}).

\end{thebibliography}

\end{document}